\documentclass[10pt,letterpaper]{article}

\usepackage[OE]{express}
\usepackage{amsmath,
amssymb,
graphicx,
color,
tikz,
pgfplots,
geometry,
citesort,
cleveref}

\usepackage{todonotes} 

\newcommand{\ket}[1]{|#1 \rangle}

\newcommand{\braket}[2]{\langle #1| #2\rangle}

\newcommand{\ncl}{n_{\text{cl}}}
\newcommand{\nco}{n_{\text{co}}}

\begin{document}

\title{Digital Waveguide Adiabatic Passage Part 1: Theory}

\author{\centering Jesse A. Vaitkus$^{1,*}$, M. J. Steel$^{2,3}$, \\ Andrew D. Greentree$^{1,4}$}

\address{\centering $^1$Chemical and Quantum Physics, School of Science, \\ RMIT University, Melbourne 3001, Australia \\
		 $^2$ ARC Centre of Excellence for Ultrahigh bandwidth Devices for Optical Systems (CUDOS) \\
		 $^3$MQ Photonics Research Centre, Department of Physics and Astronomy, \\ Macquarie University, NSW 2109, Australia \\
		 $^4$Australian Research Council Centre of Excellence for Nanoscale BioPhotonics, \\ RMIT University, Melbourne 3001, Australia}

\email{\centering $^*$jesse.vaitkus@rmit.edu.au}
\begin{abstract}
Spatial adiabatic passage represents a new way to design integrated photonic devices. In conventional adiabatic passage designs require smoothly varying waveguide separations.  Here we show modelling of adiabatic passage devices where the waveguide separation is varied digitally. Despite digitisation, our designs show robustness against variations in the input wavelength and refractive index contrast of the waveguides relative to the cladding.  This approach to spatial adiabatic passage opens new design strategies and hence the potential for new photonics devices.
\end{abstract}


\bibliographystyle{osajnl} 

\bibliography{JVlibrary}

\section{Introduction}

The continued integration of photonic devices into multi-functional chips is one of the most important drivers for the modern photonics industry \cite{PL2003}.  Integration offers increased reliability and decreased costs in the same fashion as integrated electronics enabled the computation revolution.  Notably, the robustness of integrated photonics has also enabled new generations of quantum logic devices, which are particularly sensitive to environmental fluctuations and device irregularities~\cite{OFV2009}.

Coherent adiabatic optical devices are gaining interest as they afford robust and controllable frameworks that are resilient to wavelength, realisation, or disorder in the design processes.  Here we focus on the adiabatic three-state transfer method of SAP (Spatial Adiabatic Passage) \cite{ELC+2004,GCH+2004,P2006,LDO+2007,MBA+2016} a spatial analog of the well-known STIRAP (STImulated Raman Adiabatic Passage) \cite{KTS2007}. Much work has been done to describe the properties of effective three-state systems under certain idealised conditions that neglect effects due to, amongst other things: digitization error, unequal propagation constants and  couplings beyond nearest neighbour. In the adiabatic limit the robustness against design imperfections means that many of these complications can be subsumed in the effective coupling or some loss property.

Recent work has looked into the design of adiabatic systems that use piecewise or ``digital'' control schemes instead of continuous parameter variation \cite{SMM+2007,RV2012,PA2012,VG2013,DRK2015}. It is important to stress that the concept of adiabaticity is formally inapplicable in such cases as adiabatic following is only strictly possible with continuous variation in the control parameters. Despite this, digital adiabatic passage mimics the behaviour and robustness of typical adiabatic devices. Such a design pathway opens up possibilities for systems with inherent digitisation or non-continuous devices, such as are typically found with maskless lithographic write processes.

An important technique for rapid-prototyping of integrated waveguide devices is the femtosecond laser direct write (FLDW) approach~\cite{CMH+2015,MGH+2015}. This approach uses a focused intense laser to modify the refractive index of a glass material to generate core-cladding type waveguides.  The write pattern is controlled in three dimensions, allowing highly novel devices to be achieved, including for example tritters~\cite{CMH+2015} and exotic geometries~\cite{DS2012}.  One issue with FLDW is that day-to-day reproducibility of the write power is difficult to control, which in turn affects the refractive index variation between the core and cladding.  To overcome this limitation, typically large arrays of devices with systematically varying properties are fabricated to identify the optimal device.  It is thus attractive to study device architectures that show increased robustness to such device variability.

Here we study theoretically the properties of digital adiabatic passage (DAP) applied to femtosecond laser direct-write (FLDW) integrated photonic circuits.  We consider Gaussian profile circular guides such as those which can be obtained using the femtosecond-laser direct-write (FLDW) method, which has already shown to be able to generate functioning adiabatic devices \cite{DSH+2009}, operating in the weakly guiding regime. This design has been chosen because of its structural simplicity and (semi-)analytical coupling function but neither said structure nor coupling analyticity are requirements for this method.  We generate effective tight-binding models whose couplings are verified by rigorous full-wave descriptions of these systems. These descriptions are calculated using a custom EigenMode Expansion (EME) tool \cite{NTK+2009}. We show that despite digitisation, these devices operate with high fidelity with robustness to both operating wavelength and refractive index contrast. We consider the devices here to be suitable forerunners and valid benchmarks for future novel digital systems.  We study the experimental implementation of our designs in the following paper\cite{NVC+2016}.

For any digital variation in nearest neighbor couplings it is possible to determine a compensated scheme where the lengths of the piecewise waveguide segments, which we term \emph{waveguidelets}, are varied so as to optimise the transport \cite{VG2013}. This optimisation method is compatible with any other system that can be described with  (or approximated by) a tight-binding basis \textit{inter alia} strip waveguides, (hybrid) ridge waveguides, planar waveguides, multi-core fibres, and may be useful for non-photonic systems\cite{MMA2014} opening up more new potential design opportunities.

This paper is organised as follows: we begin with an analysis of the general Hamiltonian for three-state digital adiabatic passage. Next, we use realistic writing parameters and material properties to generate the effective tight-binding model for our systems of interest.  Using these parameters we present designs for digital adiabatic passage devices, and finally we analyse some of the expected design limitations and their effects on performance, including next nearest neighbour coupling and non-uniformity in the waveguide effective refractive indices.
\begin{figure}[t]
\centering
\includegraphics[width=.2\textwidth]{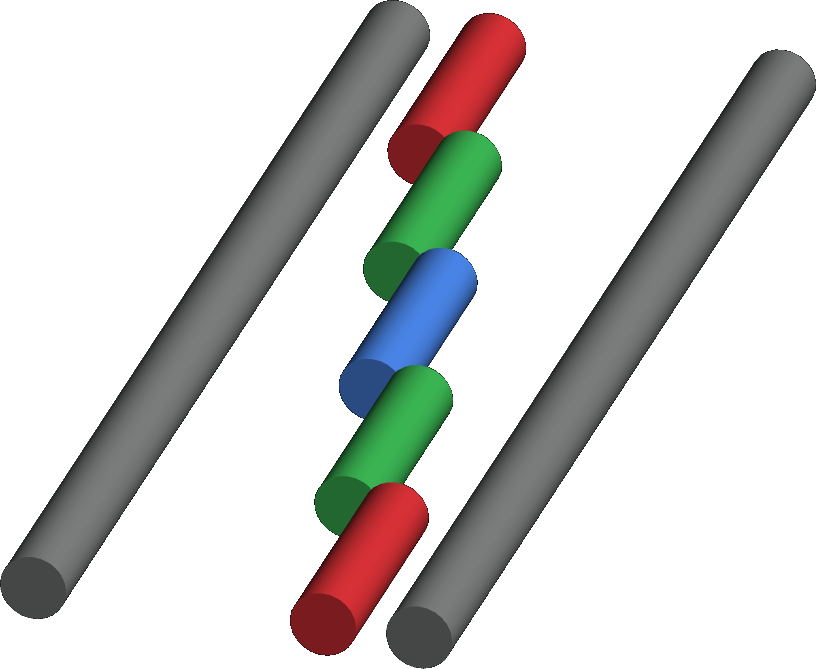}\includegraphics[width=.6 \textwidth]{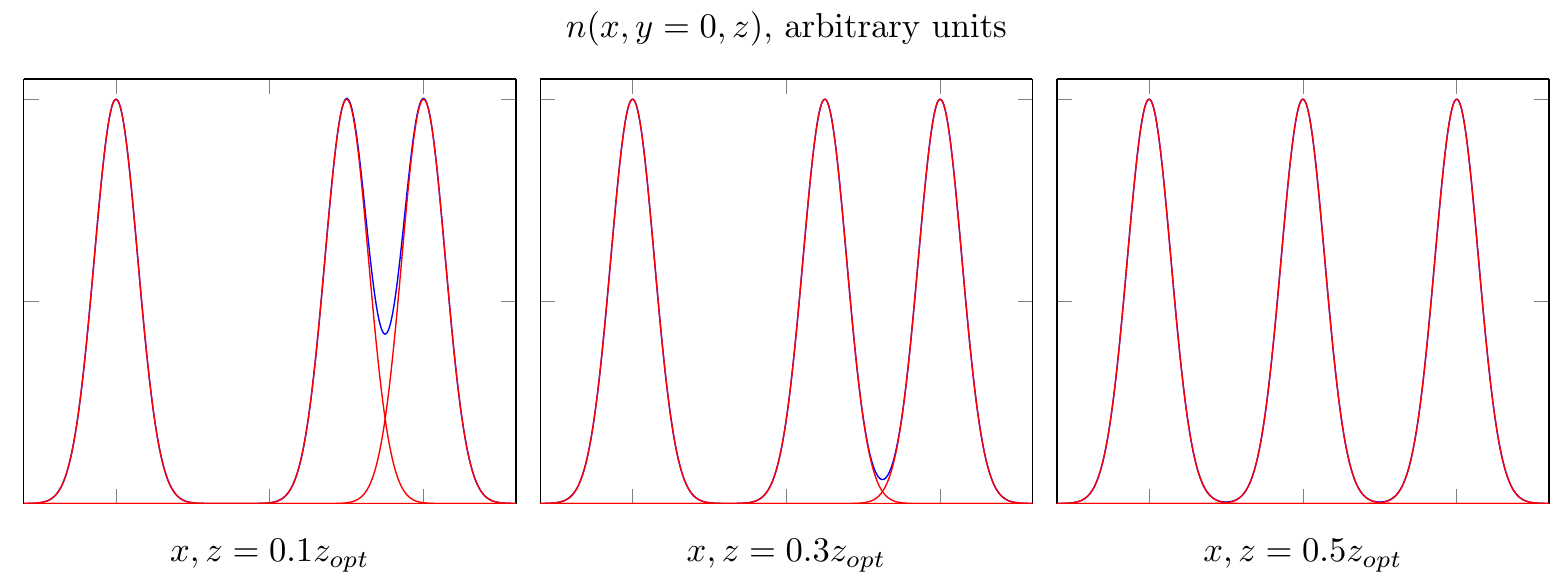}
\caption{(a) Structure for digital waveguide adiabatic passage showing the segmented waveguide with circular geometry. The counter-intuitive coupling sequence is achieved by light entering at the bottom left waveguide, and exiting via the top right, with the coupling mediated by the central waveguidelets (shown colored). Figures (b), (c) and (d) show the refractive index profiles for the red, green and blue cases from (a), demonstrating the additive nature of a continuous refractive index profile. The red lines are the refractive index of each element independently, while for the blue lines we see the sum of the refractive indices.  When the central waveguidelet is closest to one of the outer waveguides, the independent waveguide approximation breaks down.  The device is symmetric; images after the first half are not shown.}
\label{fig:examplepicture}
\end{figure}

\section{Hamiltonian}


Three-state adiabatic passage is described by the following generic Hamiltonian:
\begin{equation}
H = \left[ \begin{array}{ccc}
\beta_a & \Omega_{ab} & \Omega_{ac} \\
\Omega_{ab}^* & \beta_b & \Omega_{bc} \\
\Omega_{ac}^* & \Omega_{bc}^* & \beta_c 
\end{array} \right],
\end{equation}
where $\beta_n$ is the propagation constant for the $n^\text{th}$ waveguide and $\Omega_{nm}$ is the coupling between the $n^\text{th}$ and $m^\text{th}$ waveguides. The complex couplings in the Hamiltonian and eigenvectors are included in general for completeness but also for their relationship to STIRAP; for waveguides, the coupling is strictly real. While often idealised to have equal propagation constants and no direct next-nearest coupling, such approximations do not always hold in practical geometries.  Accordingly, here we solve the complete Hamiltonian and discuss possible loss mechanisms in the following sections.

The physics of any Hamiltonian can be described by solving for its eigenvalues and eigenvectors. To solve for the eigenvalues of a $3 \times 3$ Hamiltonian one must first solve the cubic characteristic equation, 
\begin{align}
\varepsilon_0 + \varepsilon_1 E_k + \varepsilon_2 E_k^2 + E_k^3  = 0, \label{eq:characteristic}
\end{align}
with the coefficients scaled so the term in $E_k^3$ has a unity coefficient.
Defining new variables, $Q$ and $R$:
\begin{gather}
Q \equiv \left(\frac{\varepsilon_2^2-3\varepsilon_1}{9}\right)^{1/2}, \quad 
R \equiv \frac{9\varepsilon_1\varepsilon_2-27\varepsilon_0-2\varepsilon_2^3}{54} ,
\end{gather}
then the $k^\text{th}$ eigenvalue $E_k$ is given by:
\begin{equation}
E_{k} = 2 Q \cos\left[\frac{1}{3}\arccos\left(\frac{R}{Q^3}\right) + \frac{2 \pi k}{3}\right] - \varepsilon_2/3. \label{eq:eigenvalues}
\end{equation}
Note that this solution holds if and only if all solutions to the characteristic equation (\ref{eq:approximation}) are real; this is strictly true for Hermitian matrices. Due to the arccosine, equation (\ref{eq:eigenvalues}) has an infinite number of solutions but only three such solutions are unique, and we choose $k=\lbrace 1,2,3 \rbrace$ so that the eigenvalues are linearly ordered. We use the basis $\ket{a},\ket{b},\ket{c}$ to denote the (isolated) modes of the left, middle and right waveguides respectively. The eigenvectors are then found by solving $H \ket{E_k} = E_k \ket{E_k}$ which gives
\begin{equation}
\ket{E_k} = \frac{a_k \ket{a} + b_k \ket{b} + c_k  \ket{c}}{\sqrt{|a_k|^2+|b_k|^2+|c_k|^2}}, \label{eq:eigenvectors}
\end{equation}
where
\begin{gather}
a_k = \Omega_{bc} \left[1-\frac{\Omega_{ac} (\beta_ 2 - E_k)}{\Omega_{ab} \Omega_{bc}}\right], \quad
b_k = \frac{1}{\Omega_{ab}} \left[\Omega_{ab}^* \Omega_{ac} - (\beta_a - E_k) \Omega_{bc} \right], \nonumber \\
c_k = \Omega_{ab}^* \left[ 1 - (\beta_a - E_k)(\beta_b - E_k) |\Omega_{ab}|^{-2} \right].
\end{gather}
In particular, we are interested in the spatial dark state for which $b_k=0$, which is obtained for $k=2$. For an adiabatic process, one chooses an initial state that gives significant overlap with an eigenstate and slowly vary the parameters to effect the desired outcome.  Setting $\beta_a=\beta_b=\beta_c=\beta$ and $\Omega_{ac}=0$ gives the target state
\begin{equation}
\ket{E_2} = \frac{\Omega_{bc} \ket{a} - \Omega_{ab}^* \ket{c}}{\sqrt{|\Omega_{ab}|^2+|\Omega_{bc}|^2}} ,
\end{equation}
with $E_2= \beta$. We will refer several times to the population of a given quantum state, this is defined as the modulus-squared overlap of the current state and some basis state, for instance when referring to population in the position basis:
\begin{equation}
P_i = |\braket{\psi}{i}|^2 = |\braket{i}{\psi}|^2,
\end{equation}
this can be directly related to the optical intensity in that waveguide. By initialising with all population in $\ket{a}$ (and hence all optical intensity) as well as $\Omega_{bc} \gg \Omega_{ab}$ then slowly decreasing $\Omega_{bc}$ while increasing $\Omega_{ab}$, a smooth and adiabatic passage to $\ket{c}$ is then ensured, this variation in parameters is termed the counter-intuitive sequence (the reverse, the so-called intuitive sequence). Given adiabatic following, the central guide remains unoccupied.  Hereafter all couplings are implicitly real-valued. 

\section{Tight-binding Hamiltonian}
The previous section assumed a three-state solution with arbitrarily controllable parameters.  In practice, all of the parameters are a function of the write geometry and are not completely independent. While having perfect control over device lengths (figure \ref{fig:Pseudos}) is a more direct comparison to \cite{VG2013} which focused on length-dependent effects, the more physically relevant variable for FLDW guides is that the designer likely has a high level of precision in the control of length, separation and wavelength but may have systematic imprecision in the parameters controlling magnitude of coupling and wavelength dependence. However, the DAP approach provides significant robustness, and we show that devices can be used for operation across different wavelength regimes, an advantage for practical devices.  Here we show how to calculate the tight-binding parameters from the usual waveguide modelling data. 

To account for wavelength dependent refractive index we let the cladding index $\ncl$ vary according to the Sellmeier equation of silica (SO$_2$) glass\cite{M1965} (with $\lambda$ expressed in $\mu$m):
\begin{align}
\ncl(\lambda) = \left( \frac{0.897479 \lambda ^2}{\lambda ^2-97.934}+\frac{0.696166 \lambda ^2}{\lambda ^2-4.67915 \times 10^{-3}}+\frac{0.407943 \lambda ^2}{\lambda ^2- 1.35121 \times 10^{-2}}+1 \right)^{1/2},
\end{align}
and define the refractive index difference $\delta$ and profile height parameter $\Delta$ as
\begin{align}
\delta = \nco - \ncl, \quad \Delta = \frac{\nco^2-\ncl^2}{2 \nco^2}.
\end{align}
We assume that $\Delta$ remains fixed by defining a wavelength-dependent core refractive index $\nco$ with some $\delta$ at a reference wavelength:
\begin{align}
\delta(\lambda) = \frac{\delta_{\text{ref}}}{\ncl(\lambda_{\text{ref}})}\ncl(\lambda) \quad  \implies \quad \nco(\lambda) = \left(1 + \frac{\delta_{\text{ref}}}{\ncl(\lambda_\text{ref})}\right)\ncl(\lambda).
\end{align}

To model waveguides generated by FLDW or those by some  diffusive process~\cite{CMH+2015,SL1983}, we construct the refractive index profile of the three waveguide system as the sum of local Gaussian refractive index changes:
\begin{equation}
n = \ncl + \delta \left[ \exp \left(-\frac{r_a^2}{\rho^2} \right) + \exp \left(-\frac{r_b^2}{\rho^2}\right) + \exp \left(-\frac{r_c^2}{\rho^2}\right) \right], \label{eq:wgprofile}
\end{equation}
where $\rho$ determines the $1/e$ length of the local refractive index change, $r_a,r_b$  and $r_c$ are the displacments from the peaks of the local refractive index changes that confine the modes $\ket{a}, \ket{b},\ket{c}$ respectively. For example, a linearly-varying position for the central waveguide with fixed outer guides would be given by:
\begin{align}
r_a^2 = (x+D/2)^2 +y^2, \  r_b^2 = \left[x - \left(\frac{D}{2} - d \right)\left( \frac{2 z}{z_{\max}} - 1\right)\right]^2 +y^2 , \ r_c^2=(x-D/2)^2 +y^2,
\end{align}
with $D$ the distance between the outer guides, $d$ the minimum separation between the central guide  and the other guides, and $z_{\max}$ the total length of the device in the $z$ direction. We use the couplings derived by Snyder and Love \cite{SL1983} where instead of a linearly additive profile the structure is modelled as
\begin{align}
n^2 = \nco^2 \left\lbrace 1 - 2\Delta \left[ 1 - \exp \left(-\frac{r_a^2}{\rho^2} \right) - \exp \left(-\frac{r_b^2}{\rho^2}\right) - \exp \left(-\frac{r_c^2}{\rho^2}\right)  \right] \right\rbrace  \label{eq:SLAdd}
\end{align}
Taking the square root of eq.(\ref{eq:SLAdd}) and Taylor expanding about $\nco=\ncl$ yields eq.~\eqref{eq:wgprofile} with $\delta=(\nco-\ncl)$. The maximum difference in the modelled refractive index and the one used to generate the coupling values is of order $(\nco-\ncl)^2$, which for the weakly guiding approximation $(\nco-\ncl) \sim \mathcal{O}(10^{-3})$ produces an error $\sim \mathcal{O}(10^{-6})$. To obtain the coupling values, the fundamental mode was found first by using the Gaussian approximation and minimizing the difference in propagation constants using the variational principle. The far-field electric field was then found by the so-called far-field correction, after which the coupling is found by taking the overlap of the two. Within these approximations Snyder and Love quote the fundamental mode as having error $\sim\mathcal{O}(10^{-2})$\cite{SL1983} when comparing the dimensionless fiber parameter $V = k \rho \nco \sqrt{2\Delta}$ from the numerically obtained result. We obtain the results
\begin{align} 
\Omega_{ij} =& \frac{\sqrt{2\Delta}}{\rho}  \frac{V^3(V-1)}{(V+1)^2} \exp\left[ \frac{(V-1)^2}{V+1}\right] K_0[(V-1)  R_{ij}/\rho] \label{eq:coupbessel} \\ \approx& \left(\frac{\pi \Delta }{R_{ij} \rho}\right)^{1/2}  \frac{V^3(V-1)^{1/2}}{(V+1)^2} \exp\left[(V-1)\left(\frac{V-1}{V+1} - \frac{R_{ij}}{\rho}\right)\right] \label{eq:coupexp}
\end{align}
where $R_{ij}$ is the absolute distance from one guide to another, $K_0$ is the $0^{th}$ modified Bessel function of the second kind and $V$ is the dimensionless fiber parameter.

\begin{figure}[h]
\centering
\includegraphics[width=.4\textwidth]{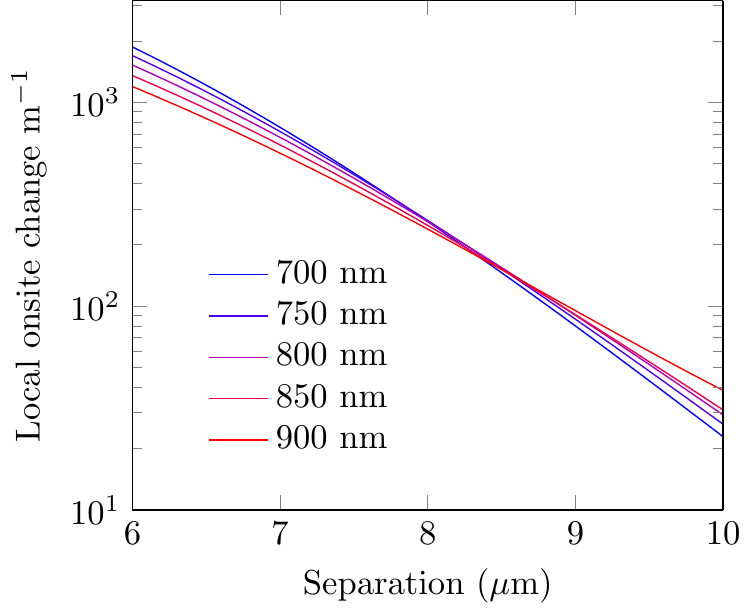}
\includegraphics[width=.4\textwidth]{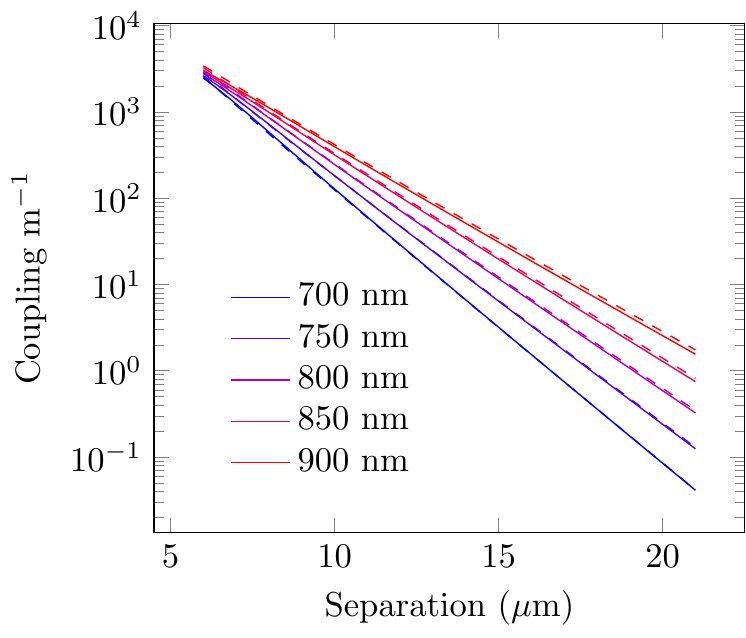}
\caption{(left) Effective change to propagation constant due to the presence of another guide. (right) Numerically (solid) and semi-analytically (dashed) obtained coupling of Gaussian index fibers.  The minimum separation is $2\rho$ so that the waveguides are clearly distinguishable. Device parameters are given in table \ref{tab:device}.}

\label{fig:coupling}
\end{figure}

Eq.~(\ref{eq:coupbessel}) gives the nearest-neighbour couplings where we have included the more commonly cited exponential approximation (\ref{eq:coupexp}) for completeness. To arrive at (\ref{eq:coupexp}) an asymptotic series of the modified Bessel function is taken. This asymptotic series leads to an over-estimate of couplings at \textit{all} separations. However, exponentially large coupling corresponds to very short distances; at such length scales the guides are no longer optically separate. Therefore both coupling functions can be used in the well-separated  regime. A comparison of the analytically and numerically obtained coupling values can be found in figure \ref{fig:coupling}.

\section{Device design}
The DAP device is realised by digitising the central waveguide of standard waveguide adiabatic passage into several parallel piecewise continuous waveguidelets. For any digital variation in $\Omega_{ab}$ and  $\Omega_{bc}$, it is possible to determine a compensated  scheme where the lengths of the waveguidelets are varied so as to optimise the transport \cite{VG2013}. The compensated scheme is robust to variations in the operating wavelength. For ideal systems with equal propagation terms or no direct next nearest neighbour ($a$--$c$) coupling, the \textit{effective} $a$--$c$ coupling rate \cite{VG2013} dictates the ideal segment length $L_{\text{opt}} = \pi/\sqrt{\Omega_{ab}^2 + \Omega_{bc}^2}$.

\begin{figure}[h!]
\centering
\includegraphics[width=.9\textwidth]{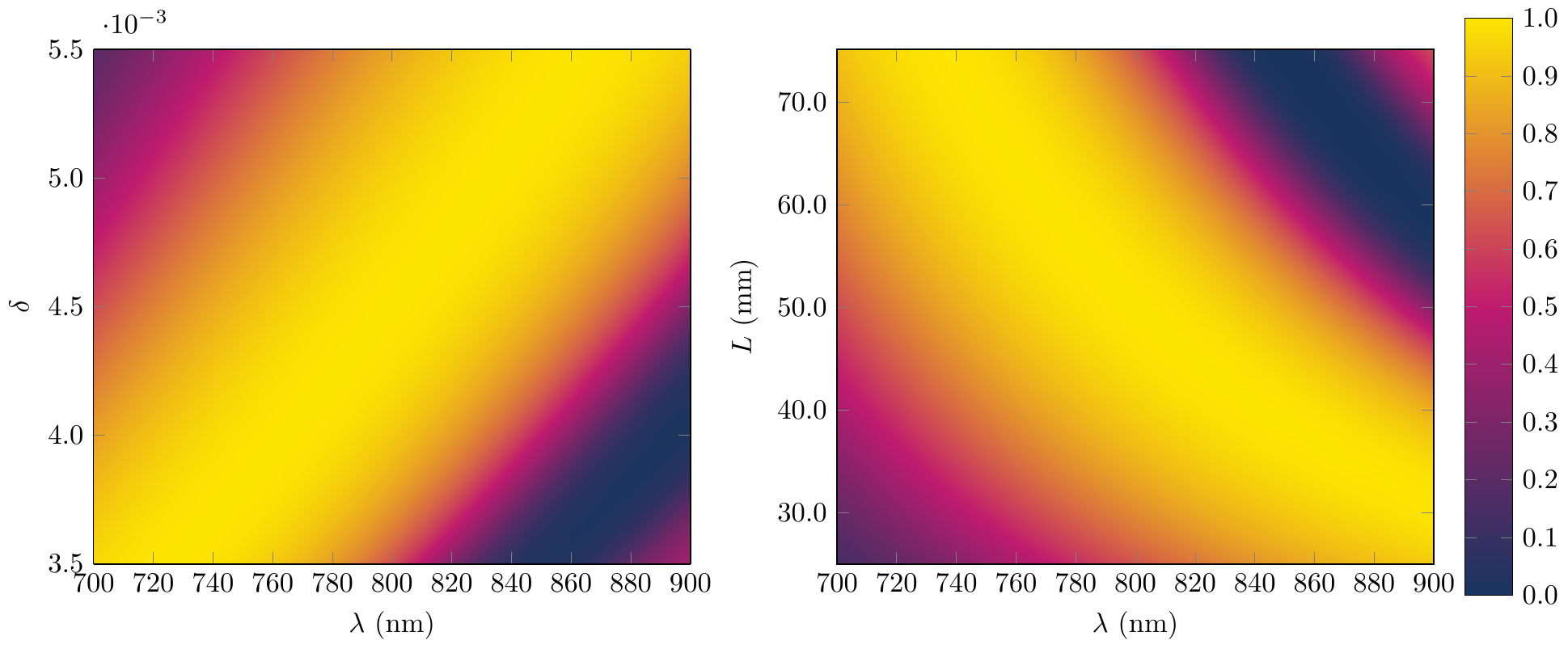}
\caption{(a) Pseudo-colour plot showing the final state population (colour axis) as a function of $\delta$ and $\lambda$.(b) Pseudo-colour plot showing final state population as a function of $\lambda$ and device length, $L$.  In both cases note the wide wavelength range over which devices provide high-fidelity transport.  The fidelity is periodic, and we have highlighted only one period here. The dark patch in the top right of the length subfigure is actually a pessimal resonance \cite{VG2013} with 90\% in the initial state despite being designed for a completely different length and wavelength.}
\label{fig:Pseudos}
\end{figure}

To most strongly demonstrate digital adiabatic passage, we separate our waveguidelets by spaces as any excitation left in the central waveguide at the end of each segment should scatter. Note that the distance between waveguidelets is irrelevant as the outer waveguides are so well separated that the exponential term in (\ref{eq:coupexp}) is orders of magnitude smaller than the smallest coupling observed and would consequently take a length several orders of magnitude larger to have any appreciable effect, therefore we set it to zero. Where there is no coupling, there is no evolution and the spaces just remove excitation rather than induce evolution within the guided modes. We also assume the propagation constants to be equal, \textit{i.e.} $\beta_a = \beta_b = \beta_c$, this equality does not always hold in general and is discussed further in the following section. As the ratio of coupling values determines the instantaneous eigenstates we choose coupling values (and hence positions) such that equal excitation is transported each step by back solving the evolution operator in \cite{VG2013}. The device parameters are shown in Table \ref{tab:device} and their resultant final state excitations in fig \ref{fig:Pseudos}.

\begin{table}[]
\centering
\caption{Device geometry and parameters used in all calculations. DAP is from $\ket{a}$ to $\ket{c}$, and the central waveguide, $\ket{b}$ is split into 5 waveguidelets, $\ket{b}_1$ to $\ket{b}_5$.  All segments are aligned at $y=0$ and $\ket{a}, \ket{b}_1, \ket{c}$ all begin at $z=0$. Segment $\ket{b}_{i+1}$ is connected to the end of segment $\ket{b}_{i}$}
\label{tab:device}
\begin{tabular}{cccccccc}\hline
Waveguidelet & $\ket{a}$ & $\ket{b}_1$ & $\ket{b}_2$ & $\ket{b}_3$ & $\ket{b}_4$ & $\ket{b}_5$ & $\ket{c}$ \\ \hline
$L_{\text{opt}} $(mm) & N/A & 7.869 & 11.270 & 11.804 & 11.270 & 7.869 & N/A \\ 
$x$ ($\mu$m) & 10.500  & -1.177 & -0.355 & 0.000 & 0.355 & 1.177 & -10.500 \\ \hline
\end{tabular}\vspace{0.5em}
\begin{tabular}{cccccc}\hline
$\rho$ & 3 $\mu$m & $\delta$ & 0.0045 & $\lambda_{\text{opt}}$ & 800 nm \\ \hline
\end{tabular}
\end{table}
The parameters in table \ref{tab:device} show a vast robustness to operating wavelength and variations in the local refractive index difference as shown in Fig. \ref{fig:Pseudos}, where the large bright regions indicate high fidelity adiabatic transport ($>\!90\%$) over a broad $100$~nm wavelength range about the optimal parameters, and indeed showing similar bandwidths away from its designed optimal range. Evidently, when one parameter deviates from its intended value, the coupling values (and hence device lengths) are no longer optimized. Despite this there are still regions of optimality. This can be explained by eq. (\ref{eq:coupexp}), where a positive increase in $\delta$, hence $\Delta$, leads to a decreased coupling, and increases in wavelength lead to increased coupling. Despite there not being a one-to-one relationship between the coupling deviations of wavelength or refractive index, the parameters shown herein are only marginally different and  result in only a low decrease in peak efficiency away from the chosen parameters ($\sim 1\%$). Indeed, a similar plot exhibiting the same features could be made for $\delta$ versus $L$.

\section{Non-dissipative physical design loss mechanisms}
As discussed earlier, any residual population in the central waveguide at the end of each segment will be scattered, reducing the overall transmission from $|a\rangle$ to $|c\rangle$ and acting as an effective source of loss.
We now discuss the possible loss mechanisms originating from reintroduction of population into $\ket{b}$ from two important effects: the next-nearest neighbour coupling and the difference in propagation constants. We stress that these are not design errors, but unavoidable consequences of realistic device geometries. That is, even when the written device has perfectly tuned $L_\text{opt}$, there will still be loss. In this section we discuss the worst case scenario, where all population is lost at each discrete step, this results in loss that increases with the number of segments, $N$. Indeed, if we connect the waveguidelets, only a certain portion of the residual population will scatter (inversely with the state overlap between sucessive waveguidelets) and thus would decrease with $N$, which is not discussed here. As stated earlier we focus on the losses originating from next-nearest neighbour coupling and a difference in the propagation constants. Each of these perturbations will shift the optimal waveguidelet length and the following derivations are derived with respect to that point; fabrication error in device length or structure is not considered. The following perturbations are cumulative, if $\Omega_{ac},\beta_{\text{diff}} \ll \Omega$; the change to the population in the central state is the sum of each contribution.

\subsection{Next nearest neighbour coupling}
\begin{figure}[h]
\centering
\includegraphics[width=0.4\textwidth]{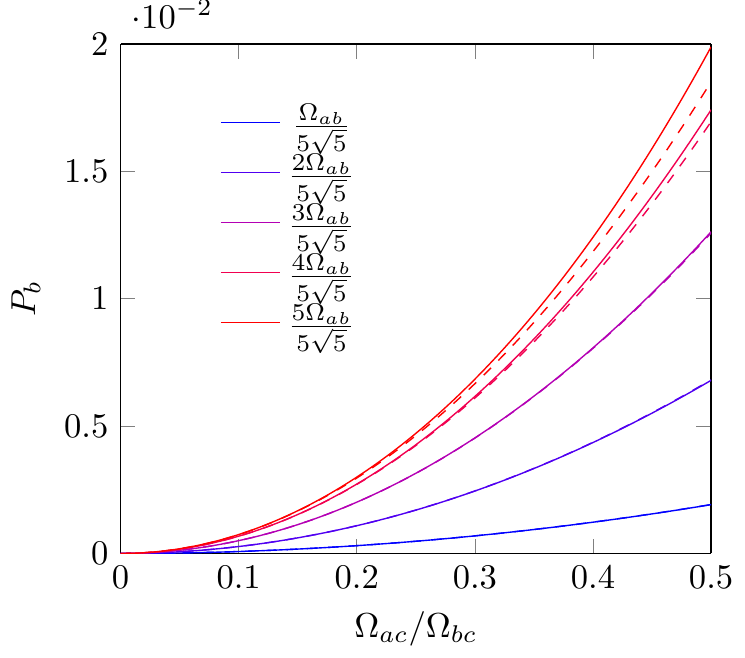}
\includegraphics[width=0.4\textwidth]{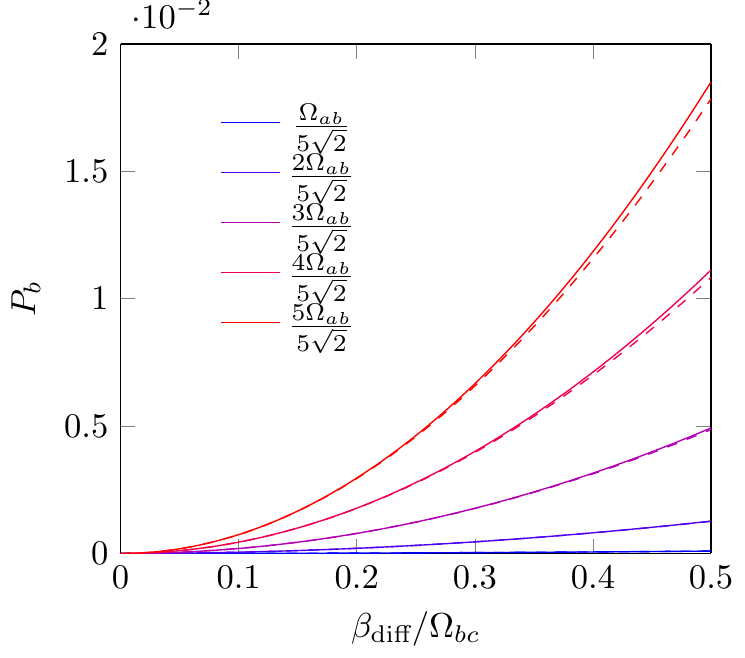}
\caption{$P_b = |\braket{E_2}{b}|^2$ as a function of scaled perturbative parameters using analytical (solid) and approximate (dashed) forms for  $\Omega_{ac}$  (left) and $\beta_{\text{diff}}$ (right). Values are symmetric with respect to $\Omega_{ab} \leftrightarrow \Omega_{bc}$, to represent data in reduced units we divide through all parameters by $\Omega_{ab}$. We also divide the perturbative parameters by $\Omega_{bc}$, as any value exceeding $\Omega_{bc}/2$ would no longer be a perturbation. The functions each have a local maximum at $\Omega_{bc}=\Omega_{ab}/\sqrt{5}$  and $\Omega_{bc}= \Omega_{ab}/\sqrt{2}$ (see \eqref{eq:approximation} and \eqref{eq:approximation2} respectively) and so values are linearly spaced up to those points. These data show that the approximations are good over a wide range of possible values with deviations strongest at the turning point.}
\label{fig:perturbs}
\end{figure}

In many device designs, next-nearest neighbour coupling (here coupling between the outer waveguides) is taken to be zero for convenience. This is typically acceptable as coupling is often negligible because three-state adiabatic passage is robust against small direct left-right coupling, and a simple  heuristic for determining when such coupling is important can be found \cite{JGC+2008}. We now consider cases where this coupling is non-zero and the implications for digital processes. We plot both approximate and analytic forms of these errors in fig \ref{fig:perturbs}. The overlap of $\ket{b}$ with $\ket{E_2}$ is analytically described by (\ref{eq:eigenvectors}), which, to first order in $\Omega_{ac}$ is:
\begin{align}
|\braket{E_2}{b}|^2 \approx \frac{\Omega_{ac}^2 (\Omega_{ab}-\Omega_{bc})^2(\Omega_{ab}+\Omega_{bc})^2}{(\Omega_{ab}^2+\Omega_{bc}^2)^{3}}. \label{eq:approximation}
\end{align}
This function is symmetric with respect to $\Omega_{ab} \leftrightarrow \Omega_{bc}$. For constant $\Omega_{ac}$ the overlap is maximal at $\Omega_{ab}=\sqrt{5} \Omega_{bc}$ or $\Omega_{bc} = \sqrt{5}\Omega_{ab}$. A comparison of the approximate and analytic form can be found in fig.~\ref{fig:perturbs}. As the loss cannot be higher than at these maxima, we introduce $\Omega=\Omega_{ab}=\sqrt{5}\Omega_{bc}$, and we use this population as an upper bound, the estimated population after $N$ steps is:
\begin{align}
1-P_{\text{Loss}}  = \left[1 - \left(\sqrt{\frac{10}{27}}\frac{\Omega_{ac}}{\Omega}\right)^2\right]^N \approx \exp\left( \frac{- 10 N \Omega_{ac}^2}{27 \Omega^2} \right). \label{eq:Om13error}
\end{align}
where the right hand side of (\ref{eq:Om13error}) uses the asymptotic form for the exponential function. This shows that the introduction of next-nearest neighbour coupling leads to an exponential increase in loss when the waveguidelets are  not connected. However, the specifics of loss accumulated will require a system-by-system analysis.

\subsection{Propagation constant mismatch}
The derivation of the tight-binding parameters relies on the waveguides and their modes being optically separable, \textit{i.e.} one can clearly distinguish when one ends and the other begins. In fig.~\ref{fig:examplepicture} we can see how designing a device by additive diffusive profiles can instigate an effective change to the propagation constants of the Hamiltonian and the independent waveguide approximation breaks down, thereby intertwining the two waveguides and their modes. The following derivation assumes that only two of neighbouring guides' propagation constants are approximately equal, and the third different, for example $\beta_b\approx\beta_c \neq \beta_a$, \textit{i.e.} the two nearest guides are strongly affecting each other but are only weakly affected by the next-nearest guide, in our design this corresponds to the waveguides confining modes $\ket{b}$ and $\ket{c}$ being close to each other, the solution also corresponds to $\beta_a\approx\beta_b \neq \beta_c$ by symmetry but the derivation itself works in general. Letting $\beta_{\text{diff}}$ be the difference between the strongly and weakly coupled guides, then the approximate on-site term for the central state is given by: 
\begin{align}
|\braket{E_2}{b}|^2 \approx \beta_{\text{diff}}^2 \frac{\Omega_{ab}^2\Omega_{bc}^2}{(\Omega_{ab}^2+\Omega_{bc}^2)^{3}}. \label{eq:approximation2}
\end{align}
When $\beta_a=\beta_c$ (regardless of $\beta_b$) the central state remains unoccupied. Unlike the previous case where $\Omega_{ac}$ was constant, the change in local refractive index depends on the nearest neighbour distance, which changes over the course of the device. To obtain an upper bound we consider that if $\beta_{\text{diff}}$ \textit{were} constant this effect would reach a maximum at $\Omega_{bc}= \Omega_{ab}/\sqrt{2}$. Therefore we set $\Omega_{ab}=\Omega$ at $\Omega_{bc}=\Omega/\sqrt{2}$, and use this centre waveguide population as an upper bound for the loss. After $N$ steps, the remaining population would be:
\begin{align}
1-P_{\text{Loss}} = \left[1 - \left(\frac{2 \beta_{\text{diff}}}{3 \sqrt{3}\Omega}\right)^2 \right]^N \approx \exp\left( \frac{- 4 N \beta_{\text{diff}}^2}{27 \Omega^2}\right),
\end{align}
where we once again have made the exponential approximation. Waveguide designs with diffusive profiles \textit{i.e.} those that locally affect each other (see figure \ref{fig:examplepicture}), will have a pronounced $\beta_{\text{diff}}$ that varies with position along the device. Adjacently coupled strip waveguides, circular cores and similar such profiles will have a lessened effect as they do not affect each others refractive indices locally. This shows that the introduction of different propagation constants leads to an exponential increase in loss when the waveguidelets are not connected. However, the specifics of loss accumulated will require a system-by-system analysis.

\section{Conclusion}
Our results indicate that digital adiabatic processes are potentially a useful new technique to be employed in the design of photonic circuits. Properties not commonly discussed such as the shift in propagation constants due to adjacent guides were also introduced and discussed. We have shown that despite digitisation, devices give high fidelity transport with broadband spectral response. This approach to spatial adiabatic passage opens new design rules and hence the potential for new/more complicated photonic devices. In the following paper \cite{NVC+2016} we show fabrication of waveguide DAP devices that confirm our predictions.

\section*{Acknowledgements}
This research was supported by the ARC Centre of Excellence for Ultrahigh bandwidth Devices for Optical Systems (Project Number CE110001018). A.D.G. acknowledges the ARC for financial support (Grant No. DP130104381).

\end{document}